\newcommand{\average}[1]{\mbox{$\langle #1 \rangle$}}
\newcommand{\ket}[1]{\mbox{$ | #1 \rangle $}}
\newcommand{\bra}[1]{\mbox{$ \langle #1 | $}}

\newcommand{\Id}{\mathds{1}}
\newcommand{\beq}{\begin{eqnarray}}
\newcommand{\eeq}{\end{eqnarray}}

\documentclass[aps,pra,twocolumn,showpacs,superscriptaddress]{revtex4-1}

\usepackage[T1]{fontenc}
\usepackage[utf8]{inputenc}
\usepackage{amsmath}
\usepackage{amsfonts}
\usepackage{amsthm}
\usepackage{amssymb}
\usepackage{subeqnarray}
\usepackage{dsfont} 
\usepackage{setspace}
\usepackage{graphics}
\usepackage{url}
\usepackage{hyperref}
\usepackage{ifthen}
\usepackage{color}
\usepackage{graphicx}
\usepackage{enumitem}
\usepackage{float}

\begin{document}

\title{Experimental comparison of tomography and self-testing in certifying entanglement}

\author{Koon Tong Goh}
\affiliation{Centre for Quantum Technologies, National University of Singapore, 3 Science Drive 2, Singapore 117543}
\affiliation{Department of Electrical \& Computer Engineering, National University of Singapore, 4 Engineering Drive 3, Singapore 117583}

\author{Chithrabhanu Perumangatt}
\affiliation{Centre for Quantum Technologies, National University of Singapore, 3 Science Drive 2, Singapore 117543}

\author{Zhi Xian Lee}
\affiliation{Department of Physics, National University of Singapore, 2 Science Drive 3, Singapore 117542}

\author{Alexander Ling}
\affiliation{Centre for Quantum Technologies, National University of Singapore, 3 Science Drive 2, Singapore 117543}
\affiliation{Department of Physics, National University of Singapore, 2 Science Drive 3, Singapore 117542}

\author{Valerio Scarani}
\affiliation{Centre for Quantum Technologies, National University of Singapore, 3 Science Drive 2, Singapore 117543}
\affiliation{Department of Physics, National University of Singapore, 2 Science Drive 3, Singapore 117542}

\begin{abstract}
We assess the quality of a source of allegedly pure two-qubit states using both standard tomography and methods inspired by device-independent self-testing. Even when the detection and locality loopholes are open, the latter methods can dispense with modelling of the system and the measurements. However, due to finite sample fluctuations, the estimated probability distribution usually does not satisfy the no-signaling conditions exactly. We implement data analysis that is robust against these fluctuations. We demonstrate a high ratio $f_s/f_t\approx 0.988$ between the fidelity estimated from self-testing and that estimated from full tomography, proving high performance of self-testing methods. 
\end{abstract}

\maketitle

\section{Introduction}

Like any physical device, quantum devices (for example, sources, transformations, and measurements) must be calibrated or certified. In this paper, we focus on \textit{certifying the properties of the state produced by a source}. The obvious technique is quantum state tomography, i.e.~the reconstruction of the density matrix. State tomography requires prior modelling, as it assumes from the start the dimension of the Hilbert space of the degree of freedom under study. It also requires modelling and calibration of the measurements.

If one is interested only in entanglement, an entanglement witness can be used instead. As well known, no measurement can detect every entangled state: a witness can only be designed with a target state in mind. Most entanglement witnesses also assume modelling of the dimension and calibration of the measurements. Famously, some don't: Bell inequalities are device-independent entanglement witnesses \cite{Werner1989,moroder2013}. There exists also a \textit{device-independent analogue of tomography, called self-testing}. Initially proved for the maximally entangled state of two qubits \cite{SW87,Popescu92,Tsirelson1993,Mayers2004,magniez2006self}, self-testing has become a rather generic and versatile tool for the a black-box certification of how close a state is to a target state $\ket{\psi}$, which must be pure and entangled, up to local isometries \cite{Yang2013,Yang2014,Bamps2015,Kaniewski2016,CGS17}. Indeed, self-testing is currently being applied to real experimental data \cite{tan2017chained,Wang2018,bancal2018device,gomez2019}. Our work contributes to this effort by applying self-testing to high quality sources, which requires the application of proper data analysis tools \cite{Lin2018}.


\section{Tomography versus self-testing: a qualitative overview}

When tasks based on Bell nonlocality are mentioned, they immediately evoke the daunting task of performing loophole-free Bell tests \cite{Hensen2015,Giustina2015,Shalm2015,Rosenfeld2017}. Closing loopholes is indeed needed to claim the label ``device-independent''. But diagnostics based on Bell nonlocality are of interest even if a loophole-free Bell test is not performed. Specifically, even if the detection and the locality loopholes are not closed, self-testing presents some advantage over standard tomography (in which fair sampling and no-signaling are routinely assumed as a consequence of the required modelling). First, it avoids the modelling assumptions on dimensions and the calibration of the measurements, which may lead to false positive \cite{rosset2012imperfect} (also see appendix). Second, it requires estimating fewer average values than tomography: for bipartite systems, three measurements on Alice and four on Bob are sufficient to assess the closeness to any pure bipartite state, of any dimension \cite{CGS17}. In summary, it is meaningful to apply self-testing tools even when the certification cannot be called device-independent in the usual sense. 

Next, we must stress that there is no free lunch: the additional assumptions give tomography some edge over self-testing. Notably, tomography can be performed on an \textit{a priori} unknown state, also of a single degree of freedom, and even in the case where (for whatever reason) the experimentalists would be targeting to produce a mixed state. By contrast, self-testing requires the target state to be known, pure and entangled: only then, it provides an estimate of the closeness of the actual state (which may be mixed of course) to the target one.

We can put this difference in a more lively narrative. The experimentalist setting up the experiment will definitely have recourse to tomography: she needs a handle over the actual degrees of freedom, and she needs potentially to scan the whole space of parameters before getting what she wants. Once the setup is up and working, she may prefer self-testing (if applicable) to convince an external referee of the quality of her source. Indeed, such a referee is unconcerned about conventional choices of bases, so he won't be bothered by the fact that self-testing is up to local isometries. But the referee may have doubts that measurements have been calibrated correctly, and will welcome a certification that does not rely on that.

This brings us to the last point of this comparison. For the experimentalist to report her results using self-testing, another condition must be met: the certification must be of \textit{comparable quality} as that obtained with tomography. If tomography yields $99\%$ fidelity with the target state, self-testing should not yield $70\%$. In this paper, we implement tools for the assessment of the self-testing fidelity on finite samples. We then apply them to experimental measurements, to characterise a source that allegedly produces pure two-qubit entangled states. We find that the self-testing fidelity can match the tomography fidelity. Thus, self-testing certification can replace tomography in reporting the quality of sources of almost pure entangled states.

\section{Theory}

\subsection{Framework}

\begin{figure}[ht]
  \includegraphics[width=\linewidth]{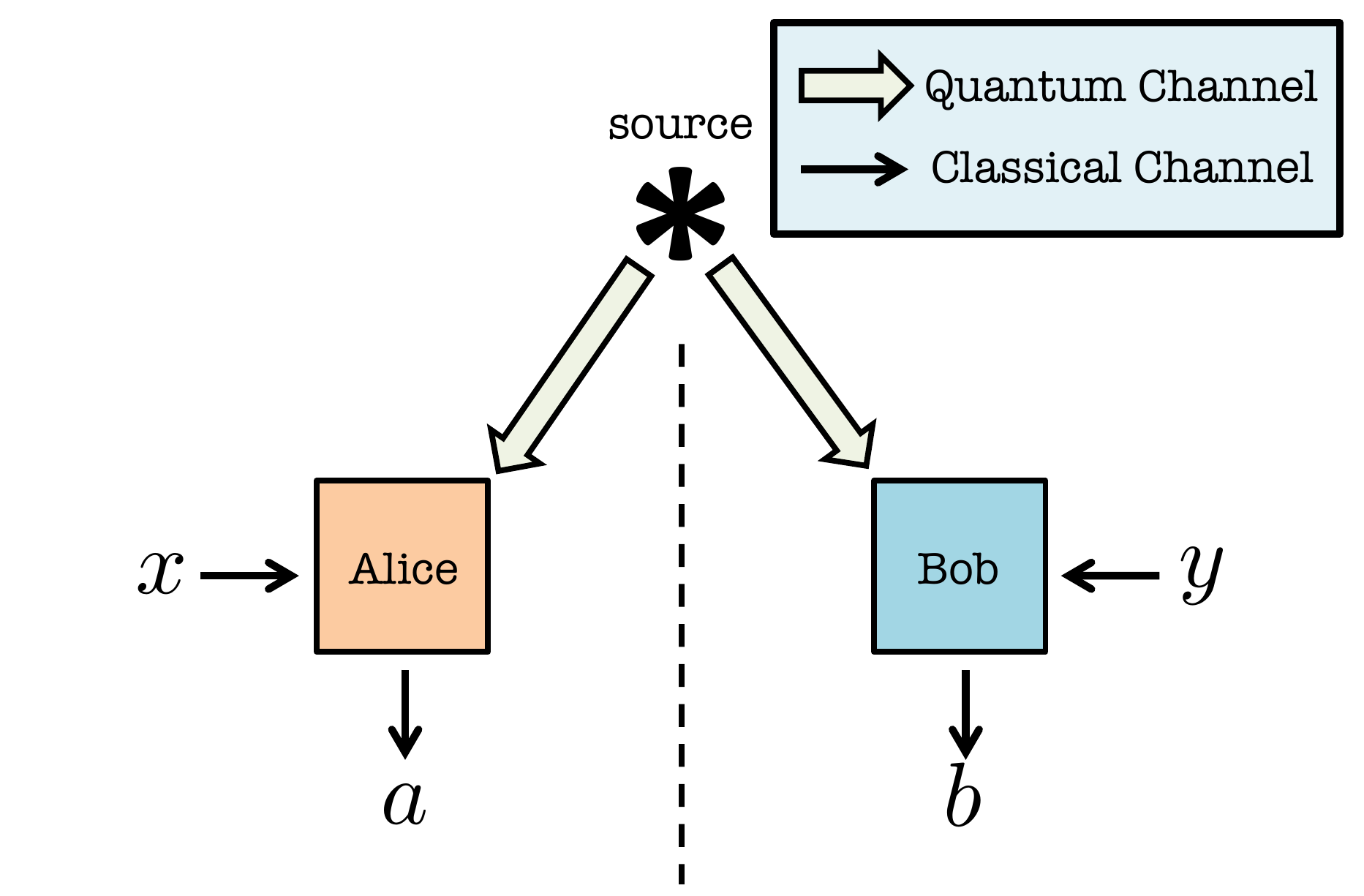}
  \caption{(Color Online) The schematic of the experiment used to certify the quality of the source.}
  \label{fig:scheme}
\end{figure}

A conceptual scheme of the setup is shown in Fig.~\ref{fig:scheme}. The \textit{source} is designed to produce, ideally, two-qubit pure entangled states. In other words, we aim at certifying how close the actual state is to one of the states described as \begin{equation}
    \ket{\psi(\theta)}=\cos{\theta}\ket{00}+\sin{\theta}\ket{11}\;,\;0<\theta\leq\frac{\pi}{4}\,, \label{idealstate}
\end{equation} up to local isometries. Each of the \textit{measurement devices}, called Alice and Bob as usual, has a classical input (denoted respectively $x$ and $y$) and a classical output (denoted respectively $a$ and $b$). Ideally, the input determines which measurement is performed, and the output is the outcomes of the measurement; we emphasize that our treatment makes no assumption on how inputs are treated or how outputs are produced. In this work, we consider binary inputs and outputs, and denote them by $x,y\in\{0,1\}$ and $a,b\in\{-1,+1\}$.

After performing several rounds of the experiment, we compute the frequencies $f(a,b,x,y)$ of each of the sixteen 4-tuples $(a,b,x,y)$; whence we estimate the conditional probabilities $P(a,b|x,y)$ through
\begin{equation}
    P(a,b|x,y)\approx\frac{f(a,b,x,y)}{\sum_{a',b'}f(a',b',x,y)}\,.
\end{equation}
It is in this estimate that we leave aside the possibility of device-independent certification. First, the probabilities are reconstructed only from events in which one detector fired on each side (thus, we assume fair sampling). Second, without arranging space-like separation between the relevant events, we assume that no side communication channel carries the information of one box's input to the other box. Under this no-signaling constraint, the sixteen $P(a,b|x,y)$ must depend on eight real parameters only: for our purposes, we take the four correlators $\average{A_xB_y}$ and the four marginals $\average{A_x}$ and $\average{B_y}$, defined by
\begin{align}
    \average{A_x} &= P(a=+1|x)-P(a=-1|x), \label{ma}\\
    \average{B_y} &= P(b=+1|y)-P(b=-1|y), \label{mb}\\
    \average{A_xB_y} &= P(a=b|x,y)-P(a\neq b|x,y). \label{corr}
\end{align}

\subsection{Self-testing of Pure Two-qubit Entangled States}

The first examples of self-testing proved that the maximal violation of the Clauser-Horne-Shimony-Holt (CHSH) inequality \cite{Clauser1969} can only by achieved, up to local isometries, by complementary measurements on a two-qubit maximally entangled state \cite{SW87,Tsirelson1993,Popescu92}. There exist a similar criterion for pure non-maximally entangled states \cite{Yang2013,Bamps2015}: for every $\alpha\in[0,2)$, the maximal quantum violation $I_{\alpha}=\sqrt{8+2\alpha^2}$ of the \textit{tilted-CHSH inequality} \cite{Acin2012} 
\begin{eqnarray}
    I_\alpha&=&\alpha\average{A_0}+\average{A_0B_0}+\average{A_0B_1}+\average{A_1B_0}-\average{A_1B_1}\nonumber\\
    &\leq& 2+\alpha \label{tCHSH}
\end{eqnarray} can only be achieved by the state \eqref{idealstate} with $\alpha=2/\sqrt{1+2\tan{}^2(2\theta)}$, measured according to
\begin{align}
    A_0=\sigma_z,\,\, B_0=\cos{\mu}\sigma_z+\sin{\mu}\sigma_x,\label{idealmeasurement}\\
    A_1=\sigma_x,\,\, B_1=\cos{\mu}\sigma_z-\sin{\mu}\sigma_x,\nonumber
\end{align}
where $\mu = \arctan{(\sin{(2\theta)})}$. 

Since our source may not be ideal, we shall need a version of self-testing that is robust against experimental imperfections. The criterion itself can be made robust \cite{Bamps2015}, but one can do better than simply checking the value of $I_\alpha$, since an estimate of all the $P(a,b|x,y)$ is available from the observed values. In this paper, we will adopt the SWAP method \cite{Bancal2015,Yang2014}, based on the Navascu\'{e}s-Pironio-Ac\'{\i}n (NPA) \cite{NPA2008} hierarchies of relaxation of the set of quantum correlations.

\subsection{Finite Sample Size Effects}

In the first report of the application of self-testing to experimental observations \cite{Wang2018}, the bound on the fidelity with the target state was estimated by plugging the observed frequencies into the expressions of suitable Bell-type inequalities. Here, we implement a previous data processing, addressing concerns that arise due to statistical fluctuations.

\begin{figure}[ht]
  \includegraphics[width=\linewidth]{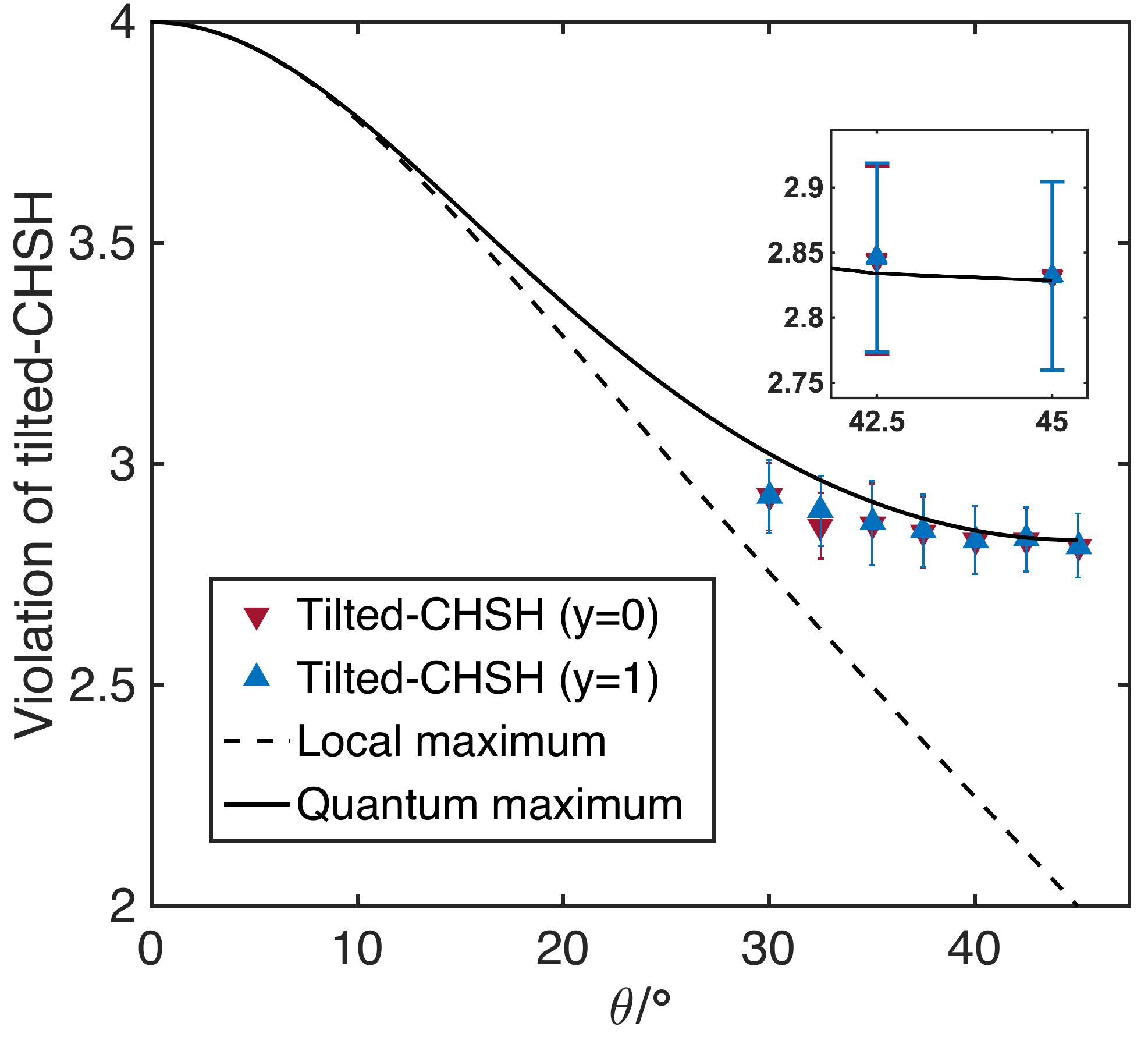}
  \caption{(Color Online) Observed violation of the tilted-CHSH inequalities for various values of $\theta$. The experimental results used for this plot consist of 500 trials per setting $(x,y)$. The error bars represent standard deviation obtained by assuming a Poisson distribution of photon counting. For the same experimental results, we compute $\average{A_0}_y:=\sum_{a,b} a P(a,b|x=0,y)$ for both $y=0,1$. This plot is an example that illustrates two concerns that call for a proper finite sample analysis. First, for $\theta=32.5^\circ$ the point estimator clearly violate the no-signaling condition, as $\average{A_0}_{y=0}\neq \average{A_0}_{y=1}$. Second, the plot in the inset are obtained from experimental results consist of 100 trials per setting $(x,y)$, which is one-fifth of that for points outside of the inset. For $\theta=42.5^\circ,45^\circ$, the sample mean exceeds the theoretical quantum maximum.}
  \label{fig:violation}
\end{figure}

The awareness of the importance of statistical fluctuations due to the finite size of the samples is rather recent even in normal tomography \cite{li2016optimal,wang2018confidence}. Notably, we mention two such concerns. The first is rather obvious: if the source is of high quality, $I_\alpha$ will be close to the quantum maximum. An estimate over few rounds may exceed that maximum, making it impossible to draw any conclusion from the point estimator. The second concern arises from the fact the probabilities inferred from the frequencies generally do not obey the no-signaling condition exactly. This is an issue because many tools in the theory of Bell nonlocality, including Bell inequalities themselves \footnote{Bell inequalities are hyperplanes bounding the set of local variables (a.k.a.~local polytope), which is obviously a subset of the no-signaling set. If we consider the whole space of probabilities, there are infinitely many hyperplanes, whose intersections with the no-signaling set define the same Bell inequality. For a given signaling probability point, one can find hyperplanes, for which the point lies on the side of the local set; and others, for which the point lies on the other side. In other words, if one takes a signaling point and plugs it into the expression of a Bell inequality, the conclusion (violation or not) may be an artefact of the choice of the Bell expression.}, can be properly used only in the no-signaling set. Using our measured results, we show an illustration of both these concerns in Fig.~\ref{fig:violation}.

We address these issues following the proposal of Lin and co-workers \cite{Lin2018}. Based on the work in \cite{Renou2017}, they devised a method to obtain a point estimator of correlation that is compatible with quantum theory from the raw observations. Since the quantum set cannot be efficiently parametrized, the point estimator is chosen as the one most compatible with the raw observations within the NPA relaxation of the quantum set of a given hierarchy level. In particular, the nearest quantum approximation (NQA$_2$) method, which uses 2-norm as a measure between correlations, can be computed efficiently using any semi-definite programming solver. We shall use this point estimator as the input for the SWAP method.

\section{Experimental Setup}
\begin{figure}[ht]
  \includegraphics[width=\linewidth]{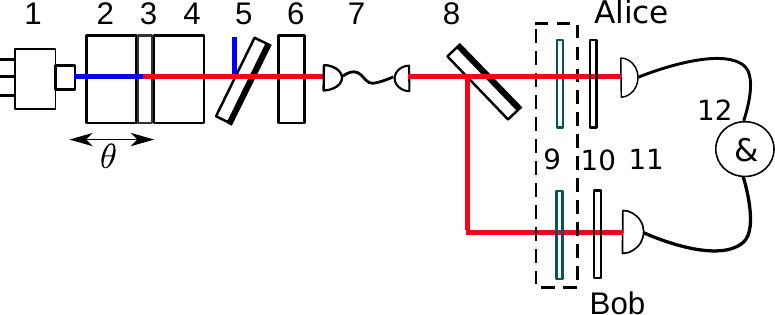}
  \caption{(Color Online) Experimental setup for the quantum state verification. 1)  405nm pump laser 2) BBO1 on a linear translational stage to change the value of $\theta$.  3) special waveplate 4) BBO2 5) dichroic mirror for removing the pump laser 6) temporal compensator (YVO$_4$) 7) Single mode fiber 8) dichroic beam splitter for splitting the signal and idler photons 9) quater wave plate (only for tomography) 10) polarizer for the projective measurements  11)  single photon detectors 12) coincidence unit for calculating $P(a,b\vert x,y)$.} 
  \label{fig:experiment}
\end{figure}

The experimental setup is sketched in Fig~\ref{fig:experiment}. Polarization entangled photon pairs are generated using spontaneous parametric down-conversion (SPDC) where a pump photon undergoes frequency conversion within a $\chi^2$  non-linear crystal to generate photon pairs of lower frequency. We have used a type-1 critically phase-matched SPDC source that produced collinear non-degenerate photon pairs from two $\beta$- barium borate (BBO) crystals whose axes are parallel \cite{Villar:18}. A pump laser of wavelength 405nm is focused to two BBO crystals of 6mm length with a special wave plate sandwiched between them. The beam waist of the focused pump is 110$\mu m$. The vertically polarized pump generates photon pairs with horizontal polarization (with state $\vert HH\rangle$) in both the crystals. The  waveplate in between the crystals rotates the polarization of the pairs produced in the first crystal ($\vert HH\rangle \longrightarrow \vert VV\rangle$) without affecting the polarization of the pump.  The wavelength dependent phase between $\vert HH\rangle$ (produced in the second crystal) and $\vert VV\rangle$  (generated at the first crystal) is compensated using a single  a-cut yttrium orthovanadate crystal (YVO$_4$) of length 3.76mm. 

The SPDC photons from both crystals are collected using a single mode fiber (SMF). The collection focus is centered at the special waveplate such that photons from both crystals are coupled with equal probabilities and generate the state $\frac{1}{\sqrt{2}}\left(\vert HH\rangle+\vert VV\rangle\right)$. In our experiment the collection beam waist is set to 60$\mu m$. If one of the BBO crystal is moved away from the waveplate, the collection to the SMF is asymmetrical, generating the state $ \cos\theta\vert HH\rangle+\sin\theta\vert VV\rangle$. The parameter $\theta$ is varied by changing the relative distance between one of the BBO crystals from the collection focus. In our experiment, BBO1 is mounted on a translation stage to generate states with $0<\theta\leq 45$. For each position of BBO1, quantum state tomography is performed to determine the exact values of $\theta$. The photons are split using a dichroic beam splitter. The polarization state of the  photons are analyzed at Alice’s and Bob’s locations to evaluate $P(a,b|x,y)$ using polarizers. Polarizer angles are decided by the value of x or y. As we are working in the non-adversarial scenario, there is no need to randomize our settings as long as we check that our measurements results have no significant drifting. In order to perform quantum state tomography, a quarter waveplate is inserted before the polarizer in Alice's and Bob's stations. The SPDC photons are then finally detected by single photon detectors. 


\section{Results and Discussion}

\begin{figure}[ht]
  \includegraphics[width=\linewidth]{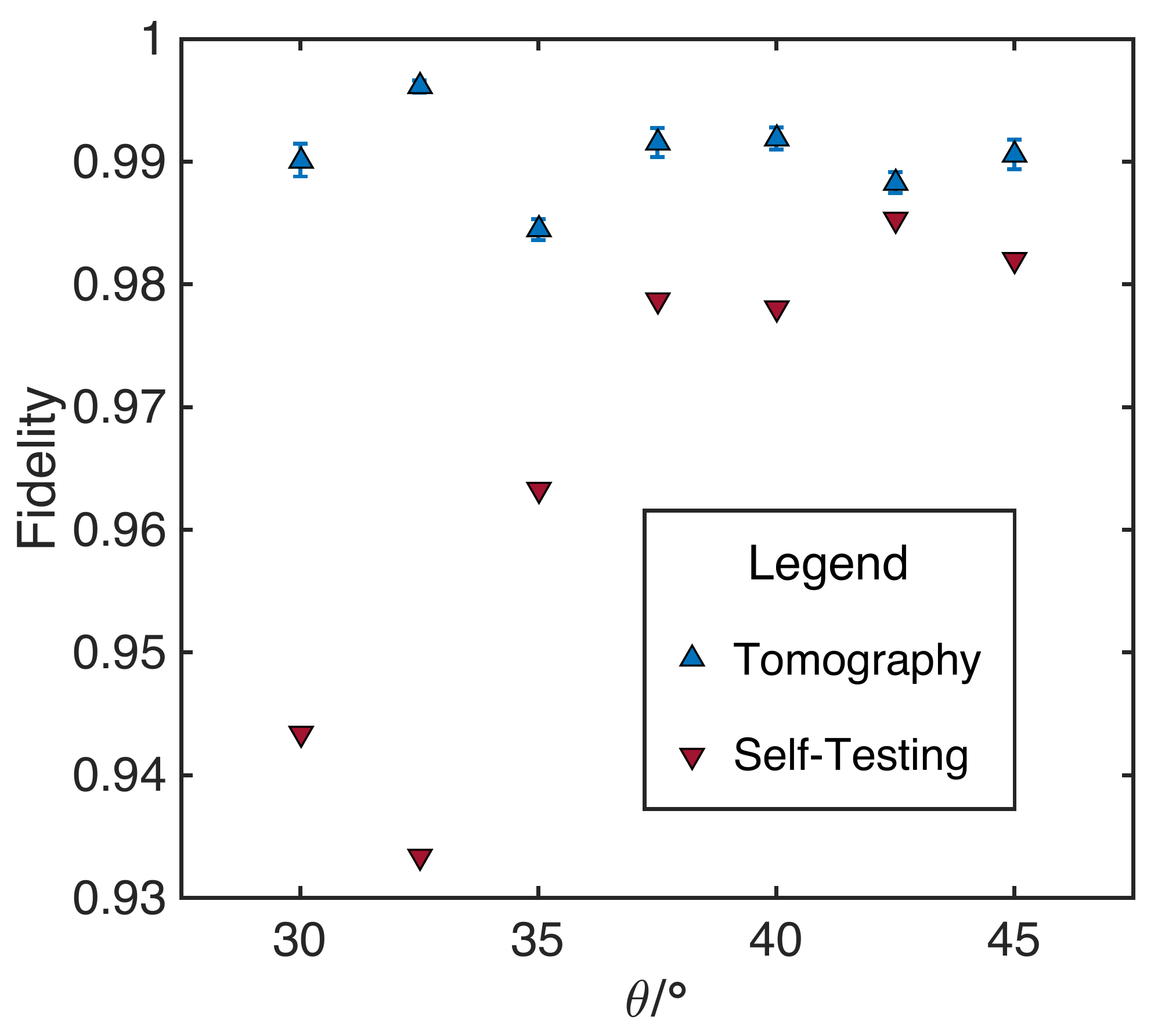}
  \caption{(Color Online) The plot of the fidelity between the measured state and the ideal state, $\ket{\psi(\theta)}$, against $\theta$. The blue upright triangular points indicate the fidelity $f_t$ obtained from quantum state tomography performed on the quantum state produced by the source in the experiment. The red inverted triangular points indicate the lower bound $f_s$ obtained by self-testing (with the measurements that maximise the tilted-CHSH inequality). These plots are obtained using a $37 \times 37$ NPA moment matrix and two $16 \times 16$ localising matrices. The values of $f_t$ and $f_s$ are listed in Table \ref{tablefid}.}
  \label{fig:fid}
\end{figure}

Using the source described earlier, we attempt to prepare the state $\ket{\psi(\theta)}$ for some pre-determined values of $\theta$ between $30^\circ$ and $45^\circ$. This range of $\theta$ is chosen due to the ample local-quantum separation of the tilted-CHSH violation with respect to the error bars (see Fig.~\ref{fig:violation}). One can extend this range by increasing the number of trials to reduce the error bars. Both tomography and self-testing were performed on the states produced in the experiment to benchmark the quality of our source. Tomography is performed using calibrated projective measurements of $\sigma_x$, $\sigma_y$ and $\sigma_z$. The fidelity between the tomographically reconstructed state and the corresponding target pure states is shown in Fig.~\ref{fig:fid}.

For self-testing, we implemented the measurements \eqref{idealmeasurement} that maximally violate the tilted-CHSH inequalities -- of course, a verifier does not need to take this piece of information into account since self-testing does not assume the calibration of measurement devices. When testing for polarization entanglement, calibration of devices usually involves the alignment of the polarization axes of polarizers or waveplates used. In our experiment, the polarizer of Alice is aligned to $0^\circ$ and $45^\circ$ (and their orthogonal angles ) while Bob's polarizer is aligned to $22.5^\circ$ and $67.5^\circ$ to achieve maximum Bell violation for $\theta =45^\circ$. However, in a practical quantum communication scenario, the exact calibration of two remote waveplates may be difficult to achieve. There can be scenarios where a) the reference axes of the polarizers or waveplates are different from one another b) the devices measure different angles due to nonlinear response (liquid crystals) or inaccurate movement (rotation stages) or c) the polarization axis of the incident photons has been shifted after the initial alignment. In such cases, tomography cannot be used to check the quality of the state, as it assumes perfect measurement settings. However, in self-testing one can always obtain a lower bound for the fidelity even if the measurement angles are off from the ideal ones. Note that with non-ideal measurement angles, the fidelity obtained by self-testing will only be an underestimation of quantum correlation present in the source.

 After the measurements are made, we apply the NQA$_2$ method on the frequencies of coincidences $f(a,b,x,y)$ to obtain the nearest set of marginals and correlators that resides in the NPA relaxation to the set of quantum correlations. We used a $37 \times 37$ NPA moment matrix detailed in the appendix. Next, we apply the SWAP method on these marginals and correlators. Here, we used the same $37 \times 37$ NPA moment matrix with additional two $16 \times 16$ localising matrices. The resulting lower bound on the fidelity with the target state is also shown in Fig.~\ref{fig:fid}.


\begin{table}[H]
    \centering
            \begin{tabular}{ |c|c|c|c|c|c|c|c| }
            \hline
             $\theta/^\circ$ & 30 & 32.5 & 35 & 37.5 & 40 & 42.5 & 45 \\ \hline
             $f_t$ & 0.990 & 0.996 & 0.984 & 0.992 & 0.992 & 0.988 & 0.991 \\  \hline
             $f_s$ & 0.943 & 0.933 & 0.963 & 0.979 & 0.978 & 0.985 & 0.982  \\ \hline
             $f_s/f_t$ & 0.953 & 0.937 & 0.979 & 0.987 & 0.986 & 0.997 &	0.991 \\ \hline
             \end{tabular}
    \caption{Fidelities obtained via tomography ($f_t$) and self-testing ($f_s$) from the experimental results.}
    \label{tablefid}
\end{table}

The fidelity obtained by tomography, denoted by $f_t$, is always higher than that obtained by self-testing, denoted by $f_s$ (see Table \ref{tablefid}). Though expected, this is not trivial: it can be taken as a validation of the modelling assumptions made for tomography. That being said, the fidelities computed from self-testing are almost identical for $\theta\geq 35^\circ$. There is nothing fundamental in this number: the range of agreement could be improved by taking larger moment matrices in the NPA hierarchy. The average ratio $f_s/f_t$ is $0.976$, but for $\theta=30.0^\circ, 32.5^\circ$, the values of $f_s/f_t$ are visibly smaller than other data points. These anomalies can be explained by the tilted-CHSH violation falling short from their maximal quantum value at these points. In Fig.~\ref{fig:violation}, we can see that the probable regions for these points excludes the value of quantum maximal violation. If one considers only the data points where the probable region of the Bell violation includes the quantum maximal violation \textit{i.e.} $\theta=35.0^\circ,37.5^\circ,40.0^\circ,42.5^\circ,45.0^\circ$, the average ratio $f_s/f_t$ is given by $0.988$. 

Our results demonstrate that even in the regime of near-maximal violation of the CHSH inequalities, self-testing (in a non-adversarial scenario) can provide a physically plausible point estimator of the bipartite entangled qubit state.
In recent work \cite{Wang2018}, similar self-testing analysis has been done for pure bipartite entangled qudits states that maximally violates the Collins-Gisin-Linden-Massar-Popescu (CGLMP) \cite{CGLMP,Kaszlikowski2000} and the Salavrakos–Augusiak–Tura–Wittek–Ac\'{i}n–Pironio (SATWAP) \cite{SATWAP} inequalities for Hilbert space dimensions up to 8. In their work, violation of CGLMP and SATWAP inequalities are obtained and used to estimate the quality of their source. However, the violations observed were far from their quantum maximal value and the analysis did not encounter the problems associated with near-maximal Bell violation using a finite sample size.

Similar can be said to any fully device-independent self-testing that was performed or will be performed in the near future. In another recent work \cite{bancal2018device}, a fully device-independent certification of the singlet state was performed and yielded a fidelity of $0.5554$ with a $99\%$ confidence. In fact, the projected near-term achievable CHSH violation by a loophole-free Bell test is given by $2.47$ \cite{Murta2018} which gives a singlet fidelity of $0.752$ using the method from \cite{Kaniewski2016}. For experimentalists who wish to check the serviceability of their entanglement source, such bounds are too pessimistic.



\section{Conclusion and Outlook}

In conclusion, we have shown that with existing quantum devices, self-testing could provide a good point estimator on the performance of a source of quantum states without assuming the characterisation of the measurements. Furthermore, this estimation is robust against false positive and requires less measurement settings as compared with quantum state tomography. This can be of great interest in practical deployment of ground or space based quantum communication systems since we can estimate the lower bound for the fidelity of the state even if the measurement devices are not calibrated.  

There is one final missing ingredient for the full solution to the problem: we could not propagate the error bars on the Bell violation and/or conditional probabilities to the error bars on the fidelity between the measured and ideal quantum states. We hope that this experimental demonstration in this paper would motivate researchers to find the full solution to the problem proposed.


\section*{Acknowledgements}
The authors would like to thank Yeong-Cherng Liang, Jean-Daniel Bancal and Yanbao Zhang for the useful discussions. This research is supported by the National Research Foundation, Prime Minister’s Office, Singapore and the Ministry of Education, Singapore under the Research Centres of Excellence programme

\bibliography{ref}
\newpage

\begin{appendix}
\section{False positive of tomography due to miscalibration of measurement}
In this section, we will illustrate a possible scenario where miscalibration of measurements would lead to false positive in quantum state tomography. 

Consider the quantum state $\ket{\phi}:=\cos{}\left(\frac{\pi}{8}\right)\ket{0}+\sin{}\left(\frac{\pi}{8}\right)\ket{1}$ which translate to the density matrix $\rho:=\ket{\phi}\bra{\phi}=\frac{\Id+\frac{\sigma_z+\sigma_x}{\sqrt{2}}}{2}$, where $\sigma_x$, $\sigma_y$ and $\sigma_z$ are the Pauli matrices.

Suppose we would like to generate $\rho$ in the lab and check the quality of our source. We could perform quantum state tomography on the produced state, denoted by $\tilde{\rho}$, and check its fidelity with $\rho$, denoted by $F(\rho,\tilde{\rho})$, which is given by:
\begin{equation}
    F(\rho,\tilde{\rho}):=\left(\text{Tr}\sqrt{\sqrt{\rho}\tilde{\rho}\sqrt{\rho}}\right)^2=\bra{\phi}\tilde{\rho}\ket{\phi}.
\end{equation}
Since the quantum state under question is a qubit state, we can write down the reconstructed state, denoted by $\rho'$, as:
\begin{equation}
    \rho':=\frac{\Id+\vec{n}\cdot\vec{\sigma}}{2},
\end{equation}
where $\vec{n}$ is the Bloch vector and $\vec{\sigma}$ is an array of Pauli matrices. Hence, one could perform quantum state tomography by making the $\sigma_x$, $\sigma_y$ and $\sigma_z$ measurements on $\tilde{\rho}$ in order to determine its Bloch vector.

In this example, suppose the prepared state $\tilde{\rho}$ is given by:
\begin{equation}
    \tilde{\rho} = p \rho+\frac{1-p}{2}\Id,
\end{equation}
for some $0\leq p\leq 1$. Hence, $\tilde{\rho}$ produces the following statistics in a tomography experiment:
\begin{align}
    \text{Tr}\left(\tilde{\rho}\sigma_z\right)&=\frac{p}{\sqrt{2}},\\
    \text{Tr}\left(\tilde{\rho}\sigma_x\right)&=\frac{p}{\sqrt{2}},\nonumber\\
    \text{Tr}\left(\tilde{\rho}\sigma_y\right)&=0.\nonumber
\end{align}
This implies that $F(\rho,\tilde{\rho})=\frac{1+p}{2}$. On the other hand, if there is a miscalibration of the measurements $\sigma_x$ and $\sigma_z$ such that the miscalibrated measurements, denoted by $\sigma'_x$ and $\sigma'_z$, are given by:
\begin{align}
    \sigma'_z&:=\cos{}(\xi)\sigma_z+\sin{}(\xi)\sigma_x,\\
    \sigma'_x&:=\cos{}(\xi)\sigma_x+\sin{}(\xi)\sigma_z\nonumber,
\end{align}
where $0\leq\xi<\frac{\pi}{2}$. Thus, the resulting statistics that is observed in the tomography experiment is given by:
\begin{align}
    \text{Tr}\left(\tilde{\rho}\sigma'_z\right)&=(\sin{}(\xi)+\cos{}(\xi))\frac{p}{\sqrt{2}},\\
    \text{Tr}\left(\tilde{\rho}\sigma'_x\right)&=(\sin{}(\xi)+\cos{}(\xi))\frac{p}{\sqrt{2}},\nonumber\\
    \text{Tr}\left(\tilde{\rho}\sigma_y\right)&=0.\nonumber
\end{align}
If one mistaken $\sigma'_z$ as $\sigma_z$ and $\sigma'_x$ as $\sigma_x$, then one would reconstuct the state $\rho'$ and conclude that $F(\rho,\rho')=\frac{1+p(\sin{}(\xi)+\cos{}(\xi))}{2}$.

Notice that when miscalibration occurs i.e. $\xi>0$, $F(\rho,\rho')>F(\rho,\tilde{\rho})$ which implies that we always get an overestimation of the fidelity between the actual state and the target state. Moreover, when $(\sin{}(\xi)+\cos{}(\xi))>\frac{1}{p}$, the estimated fidelity between the actual and target state $F(\rho,\rho')>1$, which is absurd. Thus, this analysis proves that a miscalibration of the measurements in a tomography experiment could result in false positive. 

\section{Methods for robust self-testing}
\label{apprst}
The techniques of robust self-testing that are employed in this paper will be documented in this section. In order to prove self-testing, it is sufficient to show the existence of a local isometry $\Phi(\cdot)$ such that $\Phi(\ket{\psi})=\ket{\text{junk}}\otimes\ket{\psi_\text{target}}$ where $\ket{\psi}$ is the measured quantum state, $\ket{\psi_\text{target}}$ is the target quantum state and $\ket{\text{junk}}$ can be any arbitrary quantum state. 

\begin{figure}[h]
\centering
\includegraphics[width=0.4\textwidth]{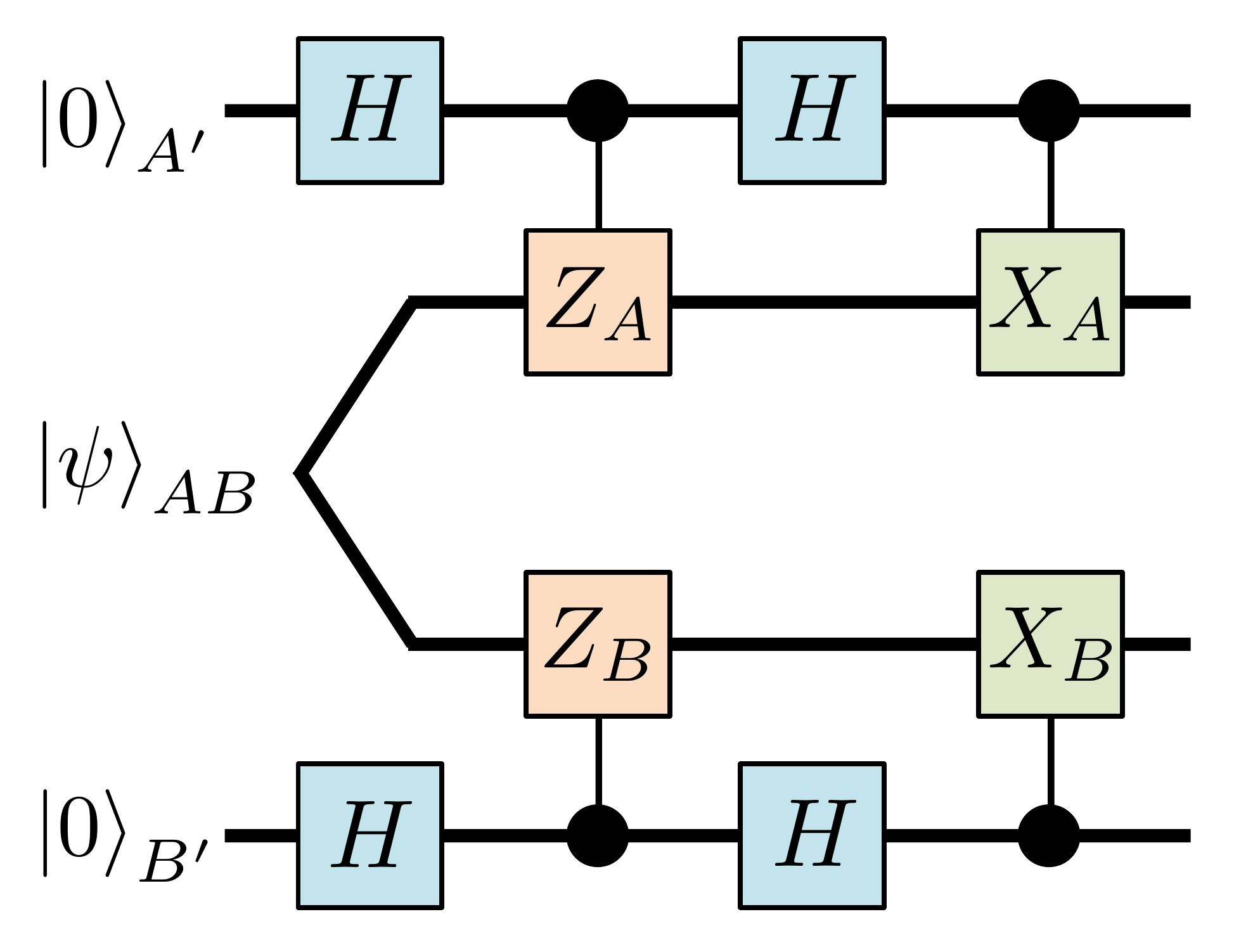}
\caption{Local isometry that is chosen for self-testing of qubits states where $H$ represents the Hadamard gate. If the operators $Z_{A/B}$ and $X_{A/B}$ correspond to the Pauli matrices $\sigma_z$ and $\sigma_x$ respectively, then the quantum circuit above corresponds to a SWAP gate between the Hilbert spaces $AB$ and $A'B'$. }
\label{fig:iso}
\end{figure}

Without loss of generality, one can pick the local isometry $\Phi(\cdot)$ to be the quantum circuit given by Figure~\ref{fig:iso}. After going through the computation of the circuit, we arrive at the following state:
\begin{align}
    \Phi(\ket{\psi}_{AB})=&\frac{1}{4}((\Id+Z_A)(\Id+Z_B)\ket{\psi}_{AB}\ket{00}_{A'B'}+\\
    &X_AX_B(\Id-Z_A)(\Id-Z_B)\ket{\psi}_{AB}\ket{11}_{A'B'}+\nonumber\\
    &X_B(\Id+Z_A)(\Id-Z_B)\ket{\psi}_{AB}\ket{01}_{A'B'}+\nonumber\\
    &X_A(\Id-Z_A)(\Id+Z_B)\ket{\psi}_{AB}\ket{10}_{A'B'})\nonumber.
\end{align}
The remaining task is to show that given the observed statistics or Bell violation, the bipartite qubits state in the Hilbert space $A'B'$, denoted by $\rho_{A'B'}$, is indeed $\ket{\psi_\text{target}}$. In the case of this paper, the target states are the pure bipartite entangled states given by $\ket{\psi(\theta)}=\cos{\theta}\ket{00}+\sin{\theta}\ket{11}$.

However, experimental results can never achieve the criteria for self-testing due to noise and error. Nonetheless, one can obtain a lower bound for the fidelity between a measured quantum state and the target quantum state, denoted by $F$, given a certain amount of deviation from the ideal statistics. Since the target states are pure, we can define the fidelity as: 

\begin{equation}
    F:=\bra{\psi(\theta)}\rho_{A'B'}\ket{\psi(\theta)}. \label{eq:fiddef}
\end{equation}

Next, in order for the unitaries $Z_{A/B}$ and $X_{A/B}$ in the local isometry to simulate the effect of $\sigma_z$ and $\sigma_x$ operators respectively, we set the operators $Z_A, X_A, \tilde{Z}_B$ and $\tilde{X}_B$ as:
\begin{align}
    Z_A&:=A_0,\\
    X_A&:=A_1,\\
    \tilde{Z}_B&:=\frac{B_0+B_1}{2\cos\mu},\\
    \tilde{X}_B&:=\frac{B_0-B_1}{2\sin\mu},
\end{align}

where $\tan\mu = \sin2\theta$. Notice that we define the operators $\tilde{Z}_B$ and $\tilde{X}_B$ with tildes as we anticipate that they are not unitary in general. Hence, inserting $\tilde{Z}_B$ and $\tilde{X}_B$ in the quantum circuit will not result in a valid local isometry. 

In order to circumvent this problem, we employ a method \cite{Yang2014}, which exploits a result from polar decomposition. For any operator $B$, there exist a decomposition such that $B=UP$, where $U$ is a unitary operator and $P$ is a positive semi-definite operator. Moreover, $P$ is unique and if $B$ is unitary, which implies $U=B$. Since, $\tilde{Z}_B$ and $\tilde{X}_B$ are Hermitian, one can show that the unitaries of their polar decomposition can be Hermitian. Hence, there exist some operators with $\pm1$ eigenvalues, $B_2$ and $B_3$, such that:
\begin{align}
    B_2(B_0+B_1)&\geq 0,\label{eq:Bunitary1}\\
    B_3(B_0-B_1)&\geq 0,\label{eq:Bunitary2}
\end{align}
and we define
\begin{align}
    Z_A=A_0\,,&\,X_A=A_1,\\
    Z_B=B_2\,,&\,X_B=B_3.
\end{align}
Using these relations, the optimisation to lower bound the fidelity, $F$, is given by:
\begin{align}
    \text{min}&\,\,F\label{eq:fidop2}\\
    \text{s.t.}&\,\,\Gamma\geq 0\nonumber\\
    &\,\,B_2(B_0+B_1)\geq 0\nonumber\\
    &\,\,B_3(B_0-B_1)\geq 0\nonumber
\end{align}
where $\Gamma$ is the moment matrix associated with the Navascu\'{e}s-Pironio-Ac\'{\i}n (NPA) \cite{NPA2008} relaxation that is compatible with the observed statistics $P(a,b|x,y)$. The moment matrix $\Gamma_{ij}:=\bra{\psi}O^\dagger_i O_j\ket{\psi}$ used in this paper employs the following set of operators $\{O_i\}_i$: $\Id$, $A_0$, $A_1$, $B_0$, $B_1$, $B_2$, $B_3$, $A_0A_1$, $A_1A_0$, $B_0B_1$, $B_1B_0$, $B_0B_2$, $B_2B_0$, $B_0B_3$, $B_3B_0$, $B_1B_2$, $B_2B_1$, $B_1B_3$, $B_3B_1$, $B_2B_3$, $B_3B_2$, $A_0B_0$, $A_0B_1$, $A_1B_0$, $A_1B_1$, $A_0B_2$, $A_0B_3$, $A_1B_2$, $A_1B_3$, $A_0A_1A_0$, $A_1A_0A_1$, $B_2B_3B_2$, $B_3B_2B_3$, $A_0B_2B_3$, $A_0B_3B_2$, $A_1B_2B_3$, $A_1B_3B_2$. Hence, the $\Gamma$ we use in this paper is a $37\times 37$ matrix.

Using the definition of $F$ in equation~\eqref{eq:fiddef} and the isometry given by Fig.~\ref{fig:iso}, we can compute $F$ to be given by:
\begin{align}
    F&=\frac{1}{4}(1+\average{A_0B_2}+\cos2\theta(\average{A_0}+\average{B_2})\\
    &+\frac{1}{2}(\cos\theta+\sin\theta)(\average{A_1B_3}+\average{A_1A_0B_3B_2}\nonumber\\
    &+\average{A_0A_1B_2B_3}+\average{A_0A_1A_0B_2B_3B_2}-\average{A_0A_1A_0B_3}\nonumber\\
    &-\average{A_0A_1B_3B_2}-\average{A_1A_0B_2B_3}\nonumber\\
    &-\average{A_1B_2B_3B_2})) \nonumber
\end{align}

However, the last two constraints of optimisation~\eqref{eq:fidop2} cannot be imposed in a numerical program. In order to impose the conditions~\eqref{eq:Bunitary1} to \eqref{eq:Bunitary2}, we use the method of matrix localization to provide a relaxation of the problem as it is a necessary condition that the localising matrix $\Gamma(B)_{ij}:=\bra{\psi}O^\dagger_i BO_j\ket{\psi}$ to be positive semi-definite if $B$ is positive semi-definite. Hence, we will perform the following optimisation:
\begin{align}
    \text{min}&\,\,F\label{eq:fidop3}\\
    \text{s.t.}&\,\,\Gamma\geq 0\nonumber\\
    &\,\,\Gamma(B_2(B_0+B_1))\geq 0\nonumber\\
    &\,\,\Gamma(B_3(B_0-B_1))\geq 0.\nonumber
\end{align}

In this paper, the set of operators $\{O_i\}_i$ used to construct the localising matrices are given by: $\Id$, $A_0$, $A_1$, $B_0$, $B_1$, $B_2$, $B_3$, $A_0B_0$, $A_0B_1$, $A_0B_2$, $A_0B_3$, $A_1B_0$, $A_1B_1$, $A_1B_2$, $A_1B_3$ and $A_0A_1$ (for $\Gamma(B_2(B_0+B_1))$), $A_0A_1A_0B_0$ (for $\Gamma(B_3(B_0-B_1))$). Hence, $\Gamma(B_2(B_0+B_1))$ and $\Gamma(B_3(B_0-B_1))$ we used are $16 \times 16$ matrices.

\begin{figure}[H]
  \includegraphics[width=\linewidth]{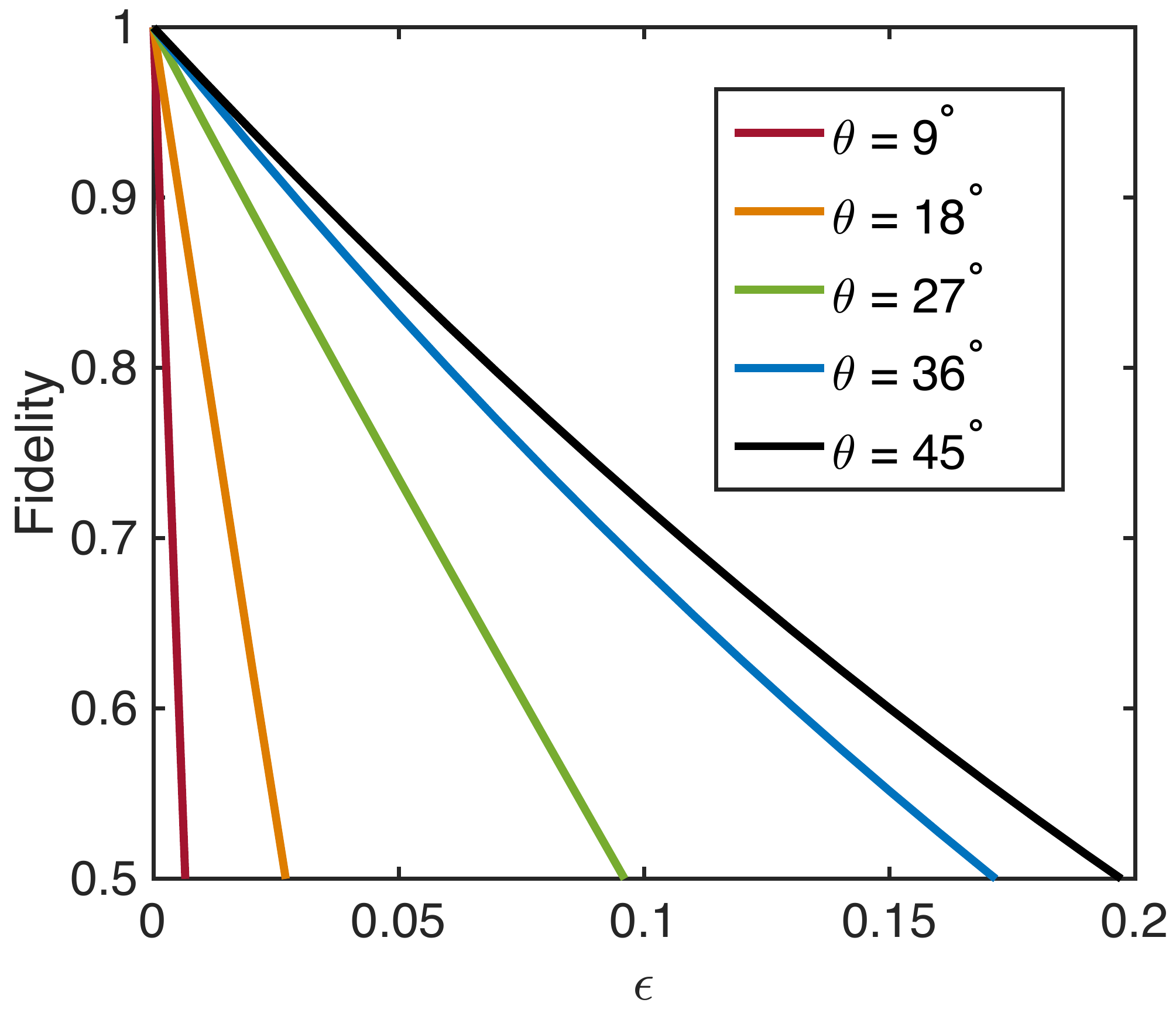}
  \caption{(Color Online) The plot of lower bounds on the fidelity between the measured state and the ideal state, $\ket{\psi(\theta)}$, against the deviation from maximal violation of the tilted-CHSH inequality denoted by $\epsilon$ (see equation~\ref{eq:epsilon}) over various values of $\theta$. These plots are obtained using a $37 \times 37$ NPA moment matrix and two $16 \times 16$ localising matrices.}
  \label{fig:robust}
\end{figure}

Performing optimisation~\eqref{eq:fidop3} for different $\theta$ over statistics with various tilted-CHSH violation gives us Fig.\ref{fig:robust}. In Fig.\ref{fig:robust}, the curves show the lower bound of the fidelity between the measured and ideal states for a given violation of the tilted-CHSH inequality. In this plot, the horizontal axis represents the deviation, denoted by $\epsilon$, from the maximal violation of the tilted-CHSH inequalities. As such, the tilted-CHSH violation is given by:
\begin{align}
    \alpha\average{A_0}+\average{A_0B_0}&+\average{A_0B_1}+\average{A_1B_0} \label{eq:epsilon}\\
    &-\average{A_1B_1}=\sqrt{8+2\alpha^2}-\epsilon.\nonumber
\end{align}

One can interpret Fig.~\ref{fig:robust} as a lookup table of the lower bound of fidelity between the measured and ideal states for a given observed Bell violation.

As the separation between the local maximum and the quantum maximum of the tilted-CHSH inequalities increase with $\theta$, the same amount of deviation from the maximal quantum violation would translate to a more drastic decrease in the lower bound on fidelity for smaller values of $\theta$ as seen in the plot.

As mentioned in the main text, the estimated conditional probabilities $P(a,b|x,y)$ may not adhere to the no-signaling constraint. In order to circumvent this problem, we employed the NQA$_2$ method \cite{Lin2018}, which essentially involves searching for the most compatible point in (the relaxation of) the set of quantum correlations with the experimental results. The NQA$_2$ method can be phrased as the following semi-definite programming problem: 

\begin{align}
    \bar{P}:=&\text{argmin}_{P}\,\,s \label{eq:regfinal}\\
    \text{s.t.}&\begin{pmatrix}
    s\Id & P-\bar{f} \\
    P^\text{T}-\bar{f}^\text{T} & s
    \end{pmatrix}\geq 0, \nonumber \\
    &\Gamma\geq 0, \nonumber
\end{align}
where $s$ is a real number, $\bar{f}$ is a vector with elements $\frac{f(a,b,x,y)}{\sum_{a',b'}f(a',b',x,y)}$, $P$ is a vector with elements $P(a,b|x,y)$ such that it is within some NPA relaxation of the set of quantum correlations i.e. $\Gamma\geq 0$ and $\bar{P}$ is a vector with its elements consisting of the regularised conditional probabilities. 

For robust self-testing of higher-dimensional pure bipartite states using conditional probabilities like the type found in \cite{CGS17}, one could adopt the SWAP method with the corresponding local isometry found therein. The solution to resulting optimization problem will provide a valid lower bound on the fidelity between the measured and target states.



\end{appendix}

\end{document}